\documentclass[prl,aps,superscriptaddress,twocolumn]{revtex4-1}
\usepackage{graphicx}
\usepackage{epstopdf}
\usepackage{amsmath}
\usepackage{ulem}
\usepackage{braket}
\usepackage[usenames,dvipsnames]{color}

\usepackage{braket}

\begin{document}

\title{Diffusion in mesoscopic lattice models of amorphous plasticity}

\author{Botond Tyukodi}
\affiliation{PMMH, ESPCI Paris, CNRS UMR 7636, Sorbonne Universit\'e, Universit\'e Paris  Diderot, PSL Research University\\
10 rue Vauquelin, 75231 Paris cedex 05, France}
\affiliation{Northeastern University, Department of Mechanical and Industrial Engineering \\ Boston, USA}
\affiliation{Babe\c{s} - Bolyai University, Department of Physics \\ Cluj-Napoca, Romania}
\author{Damien Vandembroucq}
\affiliation{PMMH, ESPCI Paris, CNRS UMR 7636, Sorbonne Universit\'e, Universit\'e Paris  Diderot, PSL Research University\\
10 rue Vauquelin, 75231 Paris cedex 05, France}
\author{Craig E Maloney}
\affiliation{Northeastern University, Department of Mechanical and Industrial Engineering \\ Boston, USA}

\pacs{--}

\begin{abstract}
We present results on tagged particle diffusion in a meso-scale
lattice model for sheared amorphous material in athermal
quasi-static conditions.  We find a short time diffusive regime and a
long time diffusive regime whose diffusion coefficients depend on
system size in dramatically different ways.  At short time, we find that
the diffusion coefficient, $D$, scales roughly linearly with system
length, $D\sim L^{1.05}$.  This short time behavior is consistent with
particle-based simulations.  The long-time diffusion coefficient
scales like $D\sim L^{1.6}$, close to previous studies which found
$D\sim L^{1.5}$.  Furthermore, we show that the near-field details of
the interaction kernel do not affect the short time behavior, but
qualitatively and dramatically affect the long time behavior,
potentially causing a saturation of the mean-squared displacement at
long times.  Our finding of a $D\sim L^{1.05}$ short time scaling
resolves a long standing puzzle about the disagreement between the
diffusion coefficient measured in particle-based models and meso-scale
lattice models of amorphous plasticity.
\end{abstract}

\date{\today}

\maketitle

Over the past few decades, the notion of local shear
transformations has been used to describe and explain the plastic flow
of amorphous solids~\cite{Bulatov1994a, Falk1998}.  A class of
mesocopic lattice models is built on this
picture~\cite{Baret2002a,TPVR-Meso12, Talamali2011a, TPRV-PRE16,
  Picard2004, Martens2011a, Nicolas2013a, Nicolas2014b, Nicolas2014c,
  Lin2014,Puosi2014a, Budrikis2013a, Budrikis2017}, (see Nicolas
  et al.~\cite{Nicolas2017} for a recent review).  In these lattice
models, the system is partitioned into local regions, and any one of
them may undergo a yielding event if loaded beyond some threshold.
These models are designed to operate at a mesoscopic scale; slightly
coarser than the particles, but not at a macroscopic scale where
continuum thermodynamical models describe phenomena such as persistent
shear localization~\cite{Falk1998,Manning2007,Bouchbinder2009}.

Avalanches of local shear transformations are observed in both
particle-scale\cite{Maloney2006,Hentschel2010,
  Salerno2012a,Salerno2013} and meso-scale models
\cite{Talamali2011a,Budrikis2013a,Lin2014b,TPRV-PRE16,Karimi-PRE17, Budrikis2017}
during slow steady shear.  The cascades are caused by the elastically
mediated redistribution of stress after a local yielding
event~\cite{Bulatov1994a,Picard2004,Eshelby1957a}.  The result is a
broad spectrum of bursts of plastic activity~\cite{Antonaglia2014} and
fractal patterns of accumulated plasticity~\cite{Sun2011}.  Similar avalanching behavior is observed in many
different dynamically critical systems~\cite{Bak1989, Bak1987,
  Tanguy-PRE98,Pindra2017, Benassi2011, Fisher1998a}.

Despite the quantitative agreement in the spectrum of avalanche sizes
and the qualitative agreement in the spatial correlations in the
plastic strain~\cite{TPVR-Meso12}, one major discrepancy between
particulate and mesoscale models has remained.  It involves the
diffusion coefficient, $D$, of the motion of tagged particles.
Lema\^{i}tre and Caroli~\cite{Lemaitre2009} argued that the spatial
correlations in the plastic strain field should give rise to a
dependence of $D$ on the system length, $L$.  In quasi-static
simulations of a Lennard-Jones glass, Maloney and Robbins
~\cite{Maloney2008} showed that $D\sim L$ out to $L\approx 1000$
particles.  In a lattice model, Martens {\it
  et. al.}~\cite{Martens2011a}, found a very different scaling with
system length for the diffusion coefficient, $D\sim L^{1.5}$.  Nicolas
and co-workers then~\cite{Nicolas2014b} showed that including the
effects of advection changes the $D\sim L^{1.5}$ scaling and suggested
including advection was necessary to obtain agreement with particulate
models.  However, very little quantitative reconciliation has been
done between the meso-scale and particulate models even for this case
of advection.

To shed light on these inconsistencies, we have performed an extensive set of simulations of a simple athermal quasi-static mesoscopic lattice model.
We find a \emph{short time} regime where $D\sim L^{1.05}$, and a
\emph{long time} regime where $D\sim L^{1.6}$.  The short time
diffusive plateau ends after a characteristic time $\Delta\gamma_*\sim
L^{-1.05}$, characterizing the strain released in a system spanning
event, which shrinks with system size.  This reconciles the $D\sim
L^1$ results of reference~\cite{Maloney2008} with the $D\sim L^{1.5}$
results of reference~\cite{Martens2011a, Nicolas2014b}.  The crossover
to the long time $D\sim L^{1.6}$ regime occurs at a size independent
strain of order unity.  At the same time, we also find consistency
with recent results of Tyukodi {\it et. al.}\cite{TPRV-PRE16}: at the
very longest times, the variance of the plastic strain field,
\emph{and, correspondingly, the mean squared particle displacement},
either continues to grow diffusively or saturates depending on whether
or not the load redistribution kernel possesses null modes.  We
further show that the kurtosis of the displacement distribution decays
with the size of the time window in the same way as in atomistic
simulations and argue that this is a generic consequence of the fact
that the displacement field is built from temporally uncorrelated shot
noise with the spatial structure of each shot being a characteristic
system spanning avalanche.

The basic approach of the lattice models goes back to Eshelby who showed that the linear elasticity problem in which a local region undergoes a shift, $\epsilon^p$, in its reference, stress-free configuration, is given by an integral convolution:
$\sigma_{\alpha\beta}(\mathbf{r})=\int
K^e_{\alpha\beta\mu\nu}(\mathbf{r}-\mathbf{r}^\prime)\epsilon^p_{\mu\nu}(\mathbf{r}^\prime)d\mathbf{r}^\prime$ where $K^e$ is the so-called Eshelby kernel.

Lattice models of amorphous plasticity then add a dynamical rule for the
evolution of $\epsilon_p$.
One of our goals in this study was to develop a simple discretization
of the Eshelby problem on a lattice which gives realistic displacements and
compatible strains near the lattice site undergoing plasticity.  
Our approach is detailed in the supplemental
material \cite{SI}, but briefly: i) we define our lattice model by
partitioning space into square domains; ii) we define the strain on
each square via a finite difference of a displacement field defined on
the vertices of the square, and iii) the response, $\sigma$, to an
increment of $\epsilon^p$ on a single square -- i.e. the Eshelby
kernel -- is expressed analytically as a Fourier series on the square
lattice and tabulated in real space for each lattice size, $L$.  This
discretization scheme is similar to that used in studies of
Martensitic transformations~\cite{Kartha,Saxena-PRB99} and to a scheme
used recently in an amorphous lattice model by
Nicolas~\cite{Nicolas2015}. At distances far from the yielding square, the shear component of this kernel gives the far-field solution of the Eshelby inclusion problem \cite{Eshelby1957a} and its shear component features a quadrupolar symmetry, i.e. in polar coordinates $K_{xyxy}^e(r, \theta) \propto \cos(4 \theta) / r^2$.

In this paper, we focus on two modes of shear with respect to the underlying lattice,
$\epsilon_{xx}=-\epsilon_{yy}=\epsilon$; $\epsilon_{xy}=0$; 
$\epsilon_{yx}=0$ which we call mode~2 and
$\epsilon_{xy}=\epsilon_{yx}=\epsilon$; $\epsilon_{xx}=0$; 
$\epsilon_{yy}=0$ which we call mode~3.  We assume the elastic
constants have the Lam\'e form so that, in either case,
$\sigma=2\mu\epsilon^e$, where $\epsilon^e$ is the elastic strain. We work in units where $2\mu=1$ so that we
can speak interchangeably of $\sigma$ or $\epsilon^e$.  Note, denoting
\emph{spatial} averages with $\langle \rangle$, that $\langle \sigma
\rangle$ and $\langle \sigma/2\mu+\epsilon^p \rangle$ would be the stress and
strain of the sample measured by a load cell.  
As we will show below, despite residual correlations at long
time in mode 2, the short time behavior of mode 2 and mode 3 in terms
of the displacement and strain statistics and the avalanche spectrum
(not studied here) is essentially indistinguishable.

We have studied different flavors of the model, characterized by different ways of introducing disorder and advancing the simulation in time.  
For the stochastic ingredients, we have studied: i) random local stress thresholds, $\sigma_y$, with uniform increments in local plastic strain,
  $\epsilon^p$, and ii) random increments in $\epsilon^p$ with uniform $\sigma_y$.
For the dynamical rule we have used: i) an extremal protocol, where the total strain $\epsilon^t = \epsilon^e + \epsilon^p$ is adjusted uniformly across the system
(it may increase or decrease) at each step so that precisely one site is at threshold~\cite{Baret2002a}, and ii) a synchronous protocol where all unstable sites are updated \emph{simultaneously} while
$\langle \epsilon^t \rangle$ is held fixed and this procedure is
iterated at the same $\langle\epsilon^t\rangle$ until all sites become
stable before $\epsilon^t$ is incremented again (also uniformly as in
the extremal protocol).  ~\footnote{Note that in the latter
  synchronous protocol avalanches are naturally produced at a discrete
  set of $\unexpanded{\langle \epsilon^t \rangle}$ with
  $\protect\langle \epsilon^t \protect\rangle$ monotonically
  increasing during the protocol, while in the former extremal
  protocol, $\protect\langle \epsilon^t \protect\rangle$ need not
  increase monotonically, and avalanches must be identified implicitly
  {\it a posteriori} by grouping sets of extremal
  events~\cite{Talamali2011a}.}
We have checked that the scaling exponents we define below do not depend on either
the stochastic model or dynamical update rule; although non-universal properties may.  The data we present here
are for the case of random $\epsilon_p$ increments with uniform
$\sigma_y$ and for the synchronous update protocol.  We choose each
local increment of $\epsilon_p$ from a uniform distribution from $0$
to $\epsilon_0$.  
For this study, we take
$\epsilon_0=1$.  
Because of the underlying
linearity, a shift in $\sigma_y$ will simply shift all loading curves
and result in precisely the same sequence of shear transformations, so
we conventionally set $\sigma_y=1$ and note that $\epsilon_0$ is the only non-trivial adjustable parameter in the
model~\cite{TPVR-Meso12,Budrikis-NatComm17}.

\begin{figure}[ht]
\begin{center}
\includegraphics[width=9cm]
{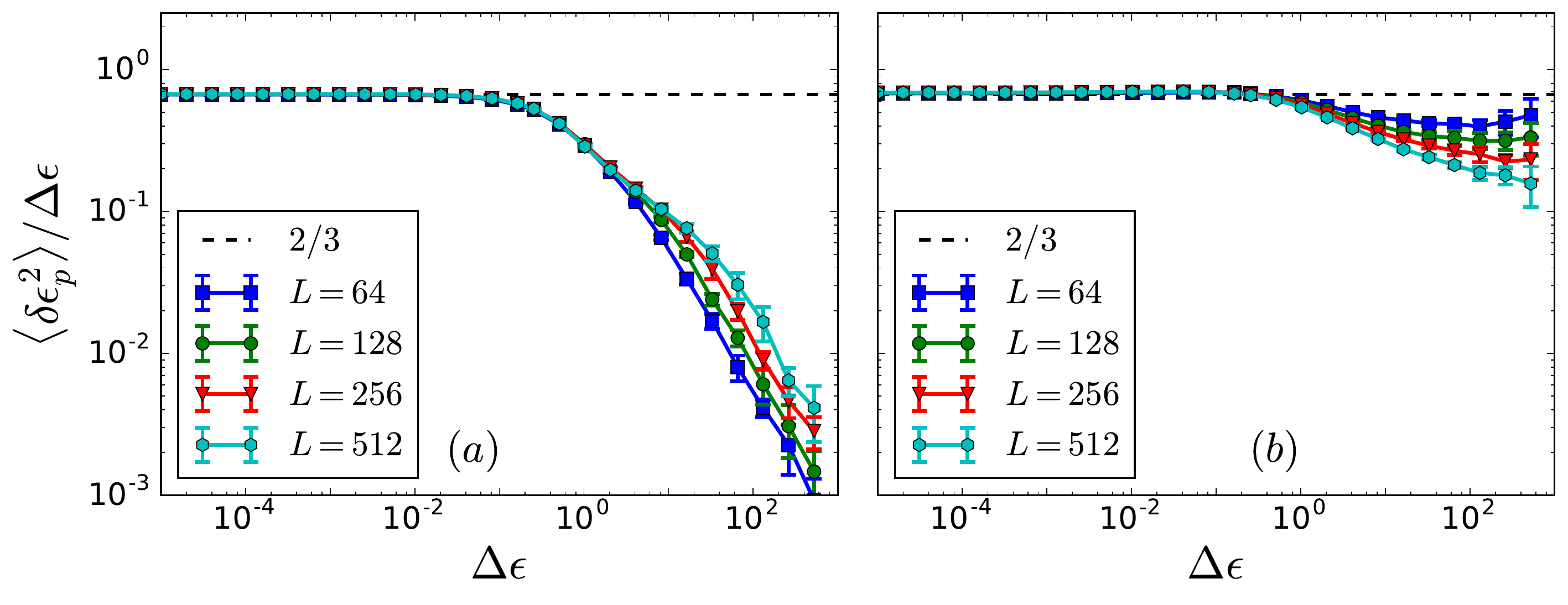}
\caption{\label{fig:mss} Variance, $\langle\delta\epsilon_p^2\rangle$,
  of the plastic strain field for a given interval of applied strain,
  $\Delta\epsilon$, scaled by $\Delta\epsilon$ for various system
  length, $L$, in a) mode~2 and b) mode~3 loading.  }
\end{center}
\end{figure}

In Figs.~\ref{fig:mss}a and \ref{fig:mss}b, we show the steady-state
variance, $\langle \delta \epsilon_p^2\rangle$, of the plastic strain
field scaled by the length of the time window, $\Delta\epsilon$, as a
function of $\Delta\epsilon$ for various system lengths, $L$, in
mode~2 (a) and mode~3 (b) loading.  For short times (small $\Delta
\epsilon$), both modes of loading show a consistent, size independent,
diffusion constant.  
We can make an {\it ab initio} estimate for the height of the plateau by assuming the probability distribution of local plastic strains is simply the uniform distribution corresponding to sites which have yielded precisely once plus a residue at zero corresponding to sites which have not yet yielded.
This {\it ab initio} estimate gives a value of $2/3$ which is in \emph{excellent} agreement with the measured plateau height.
At later times, sites will eventually undergo more than one yielding event, and this estimate will break down.

There is a fall off from the plateau, starting at
a strain of order $0.5$ (regardless of $L$) at which point each site has yielded
approximately once on average.  Beyond this fall from the plateau, the
two loading modes show dramatically different behavior.  The Mode2
curves all drop sharply.  Each curve has a shoulder feature beyond
which $\langle \delta \epsilon_p^2\rangle$ saturates and $\langle
\delta \epsilon_p^2\rangle/\Delta\epsilon\sim 1/\Delta\epsilon$.  The
shoulder extends to longer $\Delta\epsilon$ for larger $L$.  The Mode3
curves show dramatically different behavior.  After a subdiffusive
regime, the curves again approach a diffusive plateau (with a lower
diffusion constant than at short time), with the larger systems having
a lower long-time diffusion constant.  This behavior for mode~2 and
mode~3 is consistent with Tyukodi et. al.'s recent
work~\cite{TPRV-PRE16} where it was argued that the presence of null
modes in the convolution operator (and associated stress-free slip
lines) was necessary for $\langle \delta \epsilon_p^2\rangle$ to
remain diffusive.

\begin{figure}[ht]
\begin{center}
\includegraphics[width=9cm]
{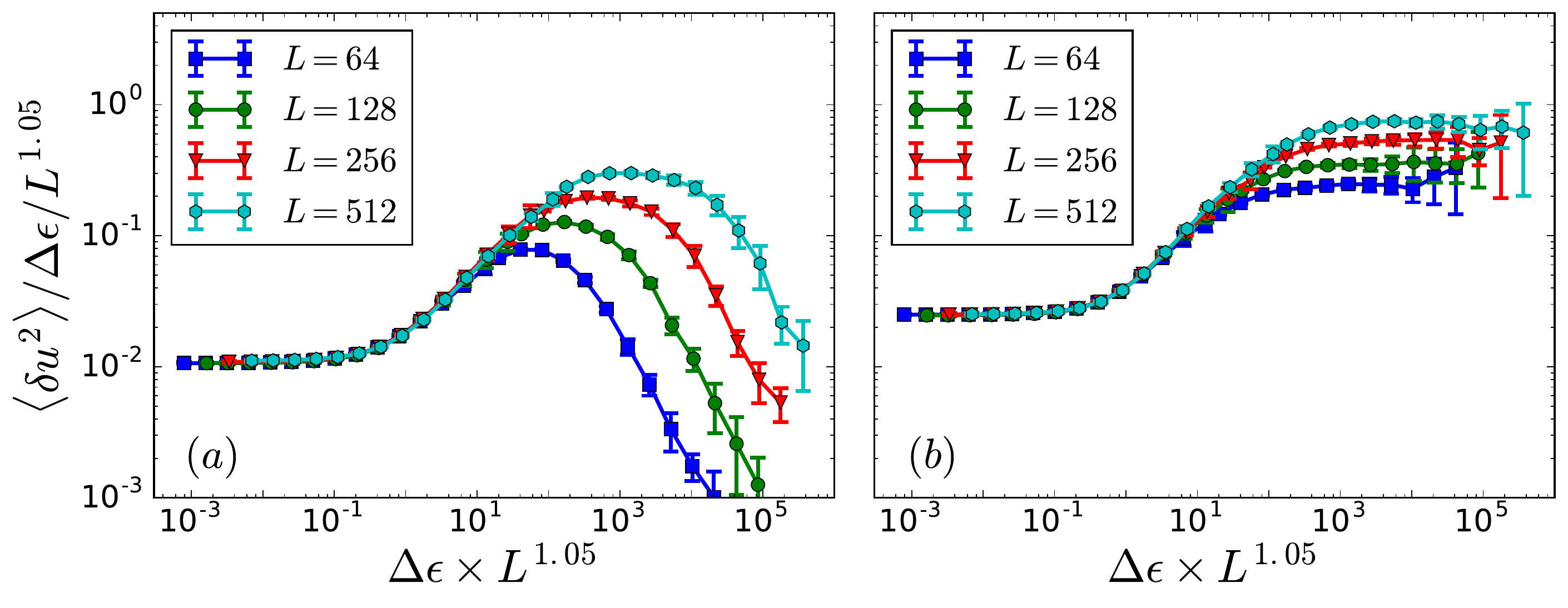}
\caption{\label{fig:diffusion} Diffusion coefficient,
  $D=\langle\delta \mathbf{u}^2\rangle/\Delta\epsilon$, of the displacement
  field scaled by $L^{1.05}$ as a function of $(\Delta\epsilon)
  L^{1.05}$ for a) mode~2 and b) mode~3.  }
\end{center}
\end{figure} 
In Fig.~\ref{fig:diffusion}, we show the diffusion coefficient,
$D=\langle \delta \mathbf{u}^2 \rangle/\Delta\epsilon $, of the
displacement field scaled by $L^{1.05}$ vs $\Delta\epsilon/L^{-1.05}$.
This rescaling collapses the data onto a short-time master curve which
has the same shape for both loading modes.  For short times, there is
a diffusive plateau.  As $\Delta\epsilon$ increases, the curve departs
upward, superdiffusively, from the plateau.  This superdiffusive
regime sets in at a characteristic time scale when:
$\Delta\epsilon_*/L^{-1.05}\approx 0.5$.

\begin{figure}[!h]
\begin{center}
\includegraphics[width=2.5cm]
{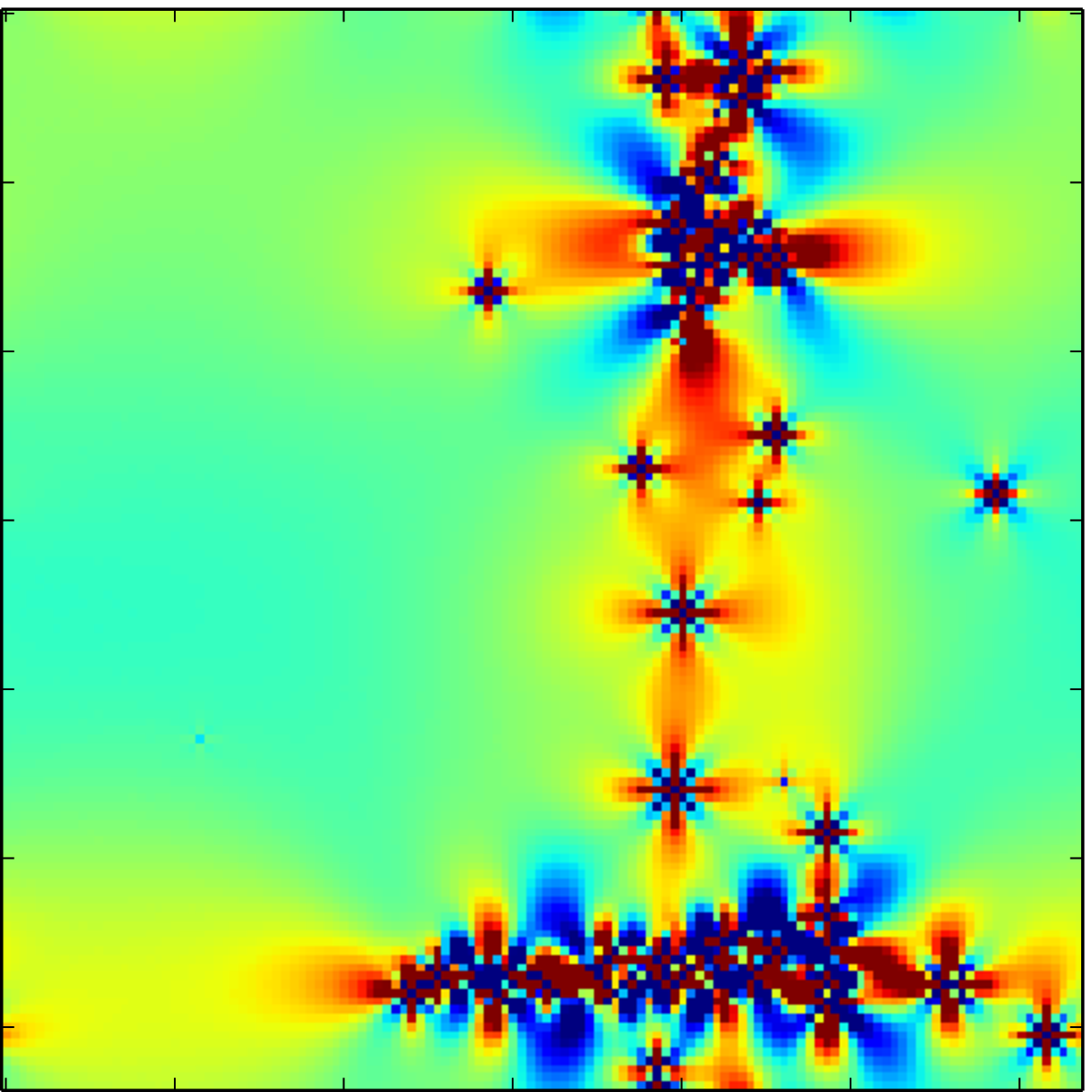}
\includegraphics[width=2.5cm]
{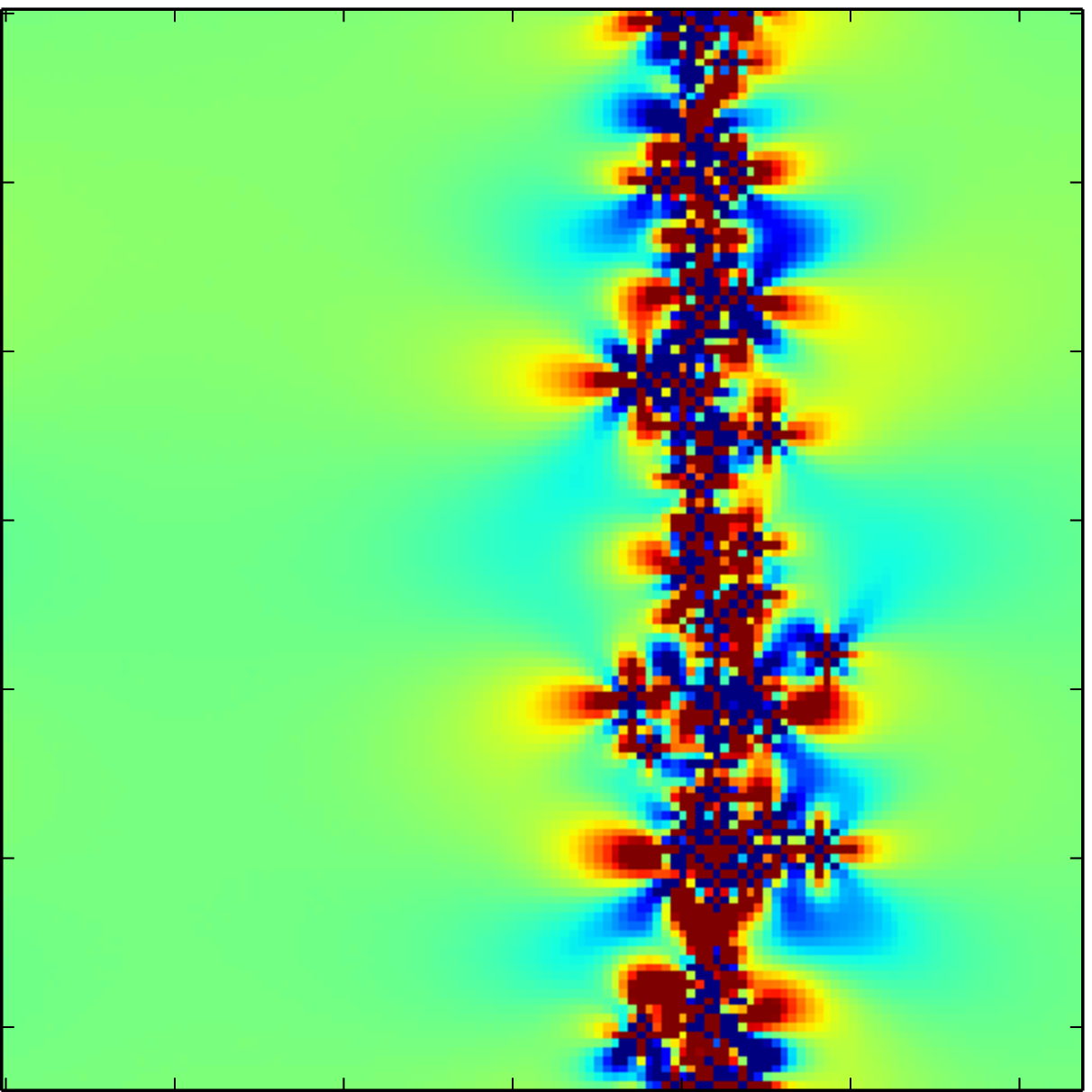}
\includegraphics[width=2.5cm]
{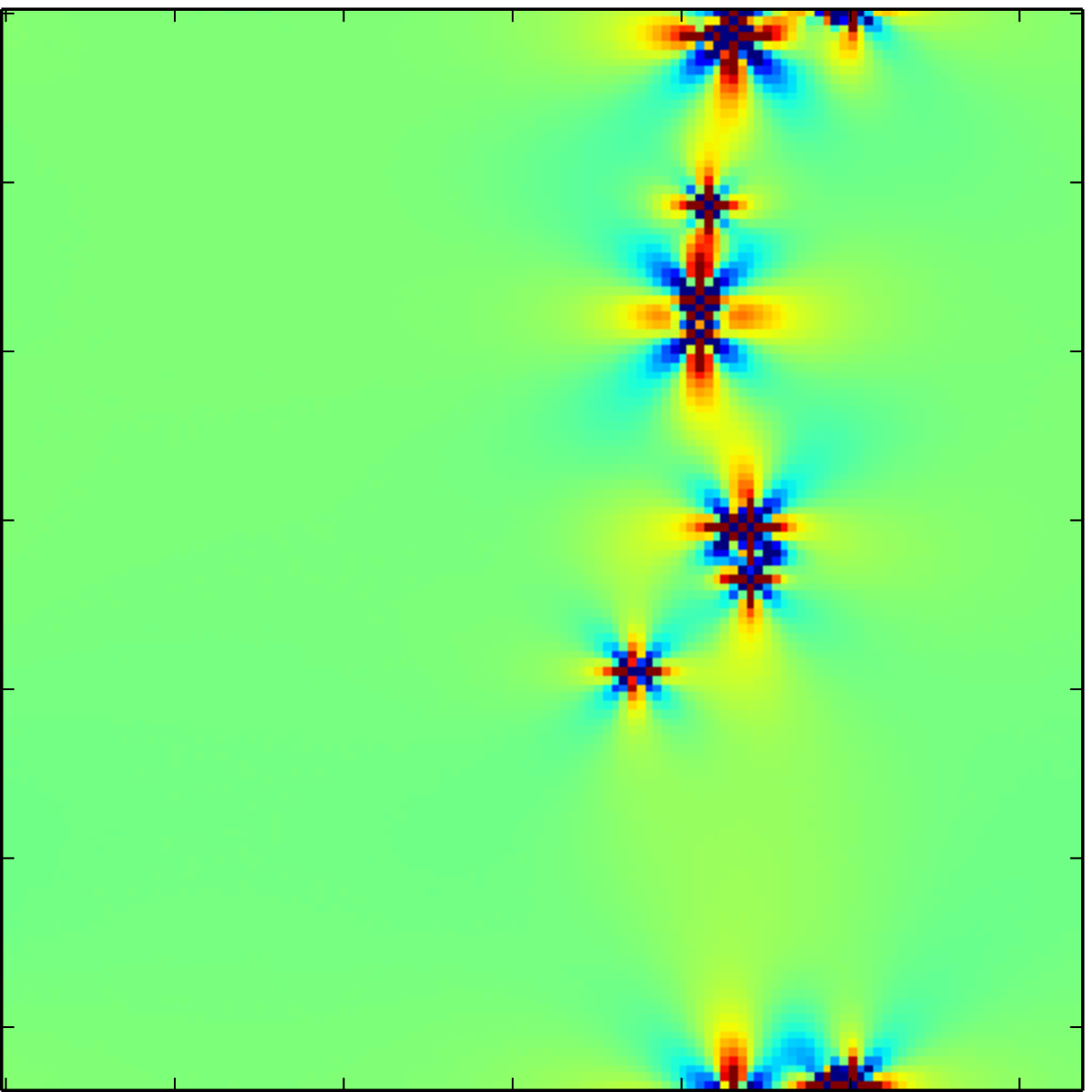} \\
\includegraphics[width=2.5cm]
{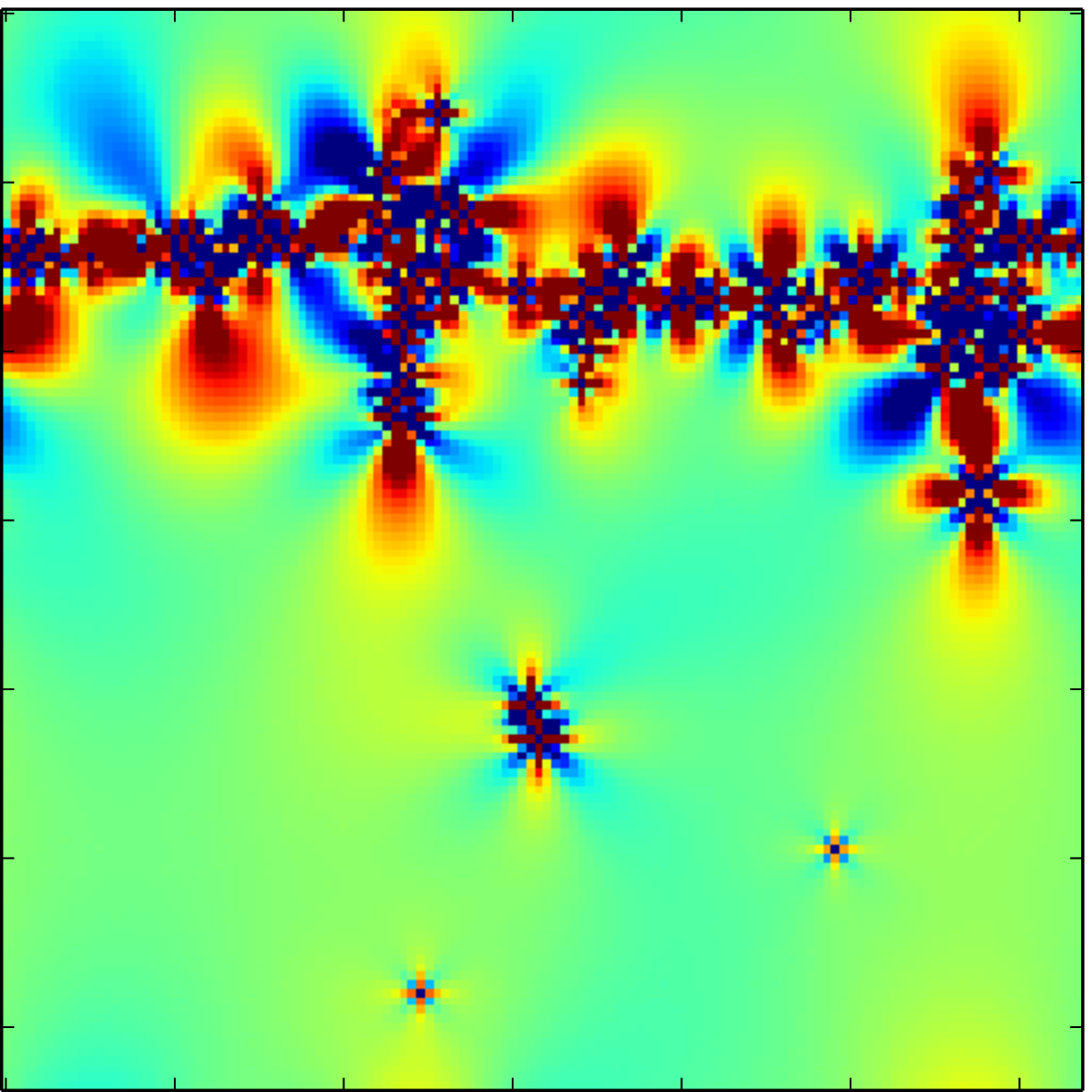}
\includegraphics[width=2.5cm]
{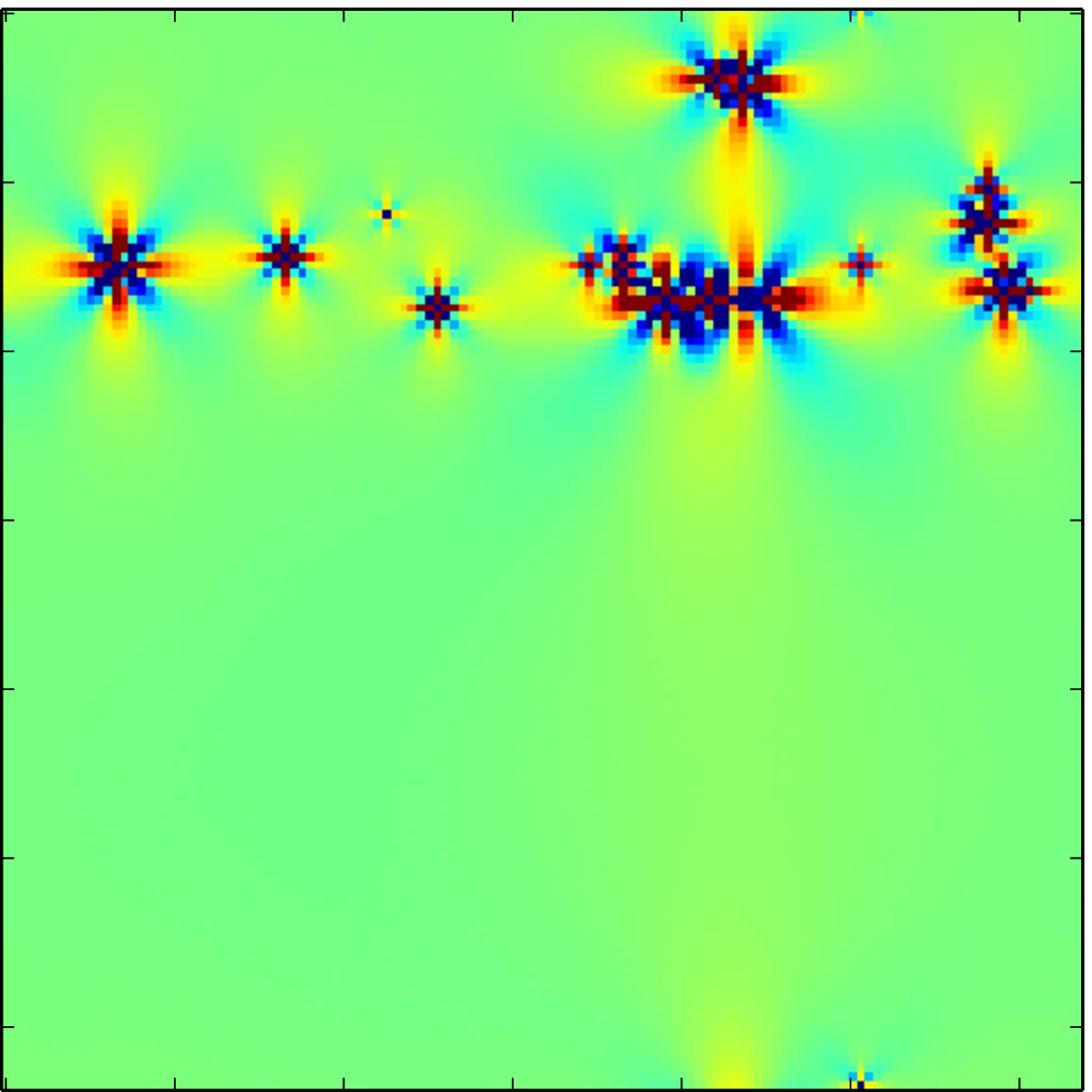}
\includegraphics[width=2.5cm]
{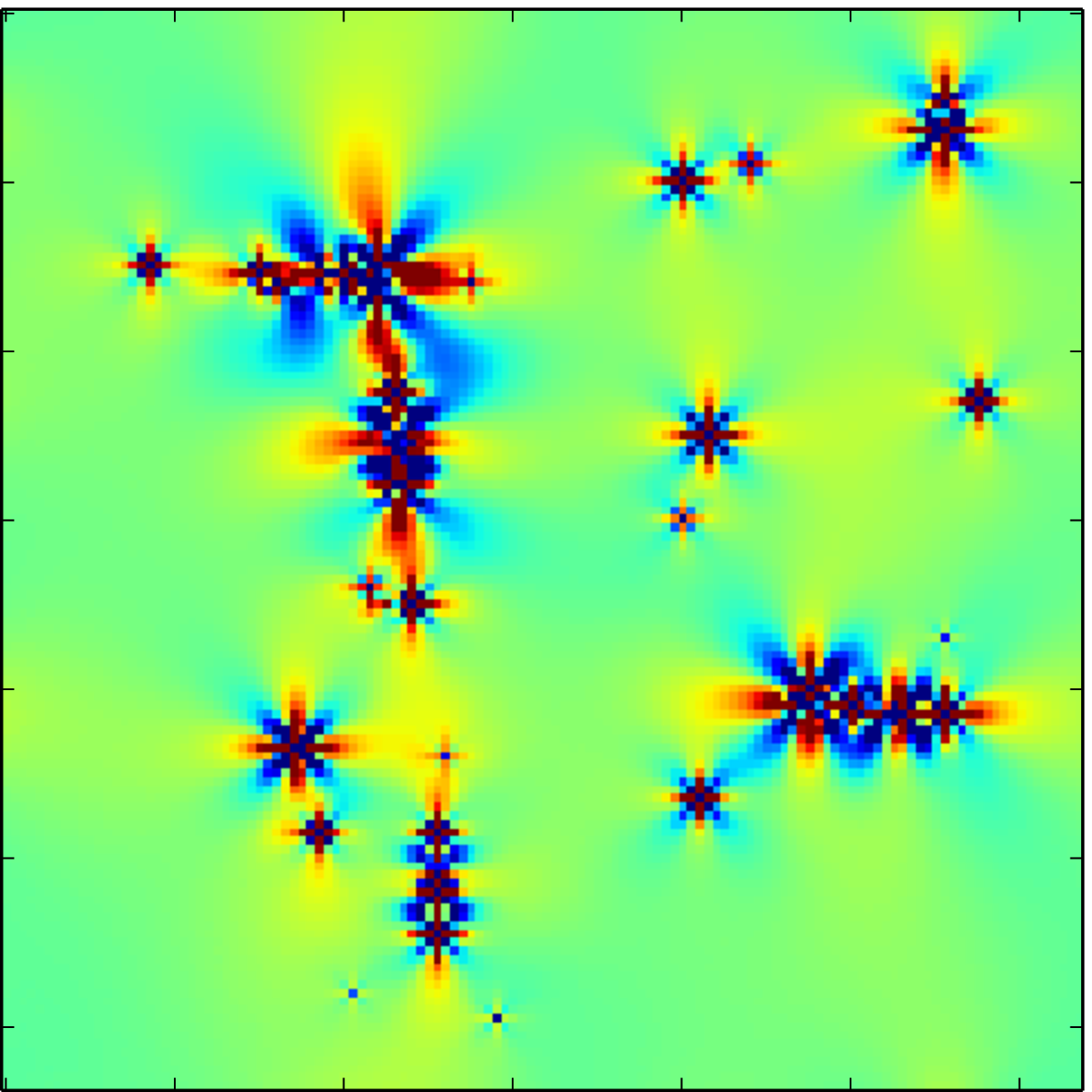} 
\caption{\label{fig:stressField} Mode~2 incremental stress field for
  several consecutive (non-overlapping) strain windows of size $\Delta
  \epsilon = 1/(2L)$ for $L=128$ such that, on average, in each window
  there are $L$ shear transformations.  This corresponds to a
  $\Delta\epsilon$ for which $D$ has just risen above the lower
  plateau in figure~\ref{fig:diffusion}.}
\end{center}
\end{figure}

To explain the scaling with $L$, we recall and generalize the
arguments of reference~\cite{Maloney2008} which were motivated by
reference~\cite{Lemaitre2009}.  In Fig.~\ref{fig:stressField}, we
plot the incremental stress field in mode~3 loading for several
consecutive non-overlapping time windows of a size corresponding to
the end of the lower plateau in figure~\ref{fig:diffusion} at the
initial stages of the superdiffusive regime.  Similar features are
observed for the other loading mode but rotated by $45$ degrees.  The
plasticity is organized into line-like features (either vertical or
horizontal) which correspond to the directions where $K^e$ is large
and positive.

Suppose the $\Delta\epsilon_p$ field for a typical time window at
short time is either zero if there has been no plasticity or composed
of a perfect line spanning the simulation cell if there has been
plasticity.  On average, each site on the line has
$\Delta\epsilon_p=\epsilon_0/2$ \footnote{The factor of $1/2$ comes
  from the uniform distribution from $0$ to $\epsilon_0$.}.  Since
there are $L/a$ such sites in the line, the whole line will globally
relieve a strain precisely equal to:
$\epsilon_{s}=a\epsilon_0/2L$ (where $a$ is the lateral size of a
  square element of the lattice).  The displacement field,
$\mathbf{u_s}$, associated with that slip line is a linear profile
with a strain equal to $a\epsilon_0/2L$ so that (assuming, for
the sake of argument, a horizontal slip line centered at $y=0$)
$u_{sx}(x,y)=2a(y-L/2)\epsilon_0/2L$.

The variance of this displacement
field is
$\langle \mathbf{u_s}^2\rangle = a^2\varepsilon_0^2/12 $
which is independent of $L$.
The rate at which these slip lines occur per unit strain,
$N/\Delta\epsilon$, has to be precisely enough so that, on average,
$\Delta\langle\epsilon_p\rangle=\Delta \epsilon$ so
$N=\Delta\epsilon/\epsilon_s=2(L/a) (\Delta\epsilon/\epsilon_0)$.  If we
are in a short time regime so that at most one of these slip lines has
formed, then we have:
$\langle \delta \mathbf{u}^2\rangle/\Delta\epsilon=N \langle
\mathbf{u_s}^2\rangle / \Delta\epsilon =(L/6)\epsilon_0 a.$
So the simpleminded picture of
elementary \emph{lines} predicts a short-time characteristic strain,
$\epsilon_s=\epsilon_0/2L$ and a short-time $D=(L/6)\epsilon_0 a^2$ .

In Fig.~\ref{fig:diffusion}, we see that the $\epsilon_* \sim L^{-1},
D\sim L^1$ scaling is only approximately correct and that scaling
$\epsilon$ by $L^{-1.05}$ and $D$ by $L^{1.05}$ gives a better quality
data collapse for both mode~2 and mode~3 loading.
We can explain this
by slightly generalizing the argument above.  If we imagine the
short-time windows contain either no plasticity or a characteristic
elementary event, then we still have that $\langle
\mathbf{u}^2\rangle/\Delta\epsilon=N \langle \mathbf{u_s}^2\rangle /
\Delta\epsilon$ where $N/\Delta\epsilon$ is still the rate of events
and $\langle \mathbf{u_s}^2\rangle$ is still the variance of a
characteristic event.  And we still must have balance between applied
strain and plastic strain so that $N=\Delta\epsilon/\epsilon_s$; but
now $\epsilon_s$ is the characteristic strain associated with an
arbitrary characteristic event more general than a straight line:
$\epsilon_s=n_s(a/L)^2\epsilon_0/2$ where $n_s$ is the number of sites
involved in one of the elementary events.  For lines, $n_s= (L/a)^1$,
while we generalize and let $n_s=A (L/a)^\alpha$ for fractal objects.  So
for the characteristic strain associated with an elementary line-like
object, we have $\epsilon_s=A(L/a)^{\alpha-2}\epsilon_0/2$.  And, finally,
for the diffusion, we have
$\langle\mathbf{u}^2\rangle/\Delta\epsilon=2 (L/a)^{2-\alpha}\langle
\mathbf{u_s}^2\rangle/(A\epsilon_0)$.
We must \emph{assume} that the elementary events produce displacement
fields whose variance is independent of $L$, but given that
assumption, we see that $D\sim L^{2-\alpha}$ and $\epsilon_s\sim
L^{\alpha-2}$.  From our scaling collapse, we conclude that
$2-\alpha=1.05$ which would correspond to a fractal dimension of
$\alpha=0.95$. 

 \begin{figure}[ht]
\begin{center}
\includegraphics[width=9cm]{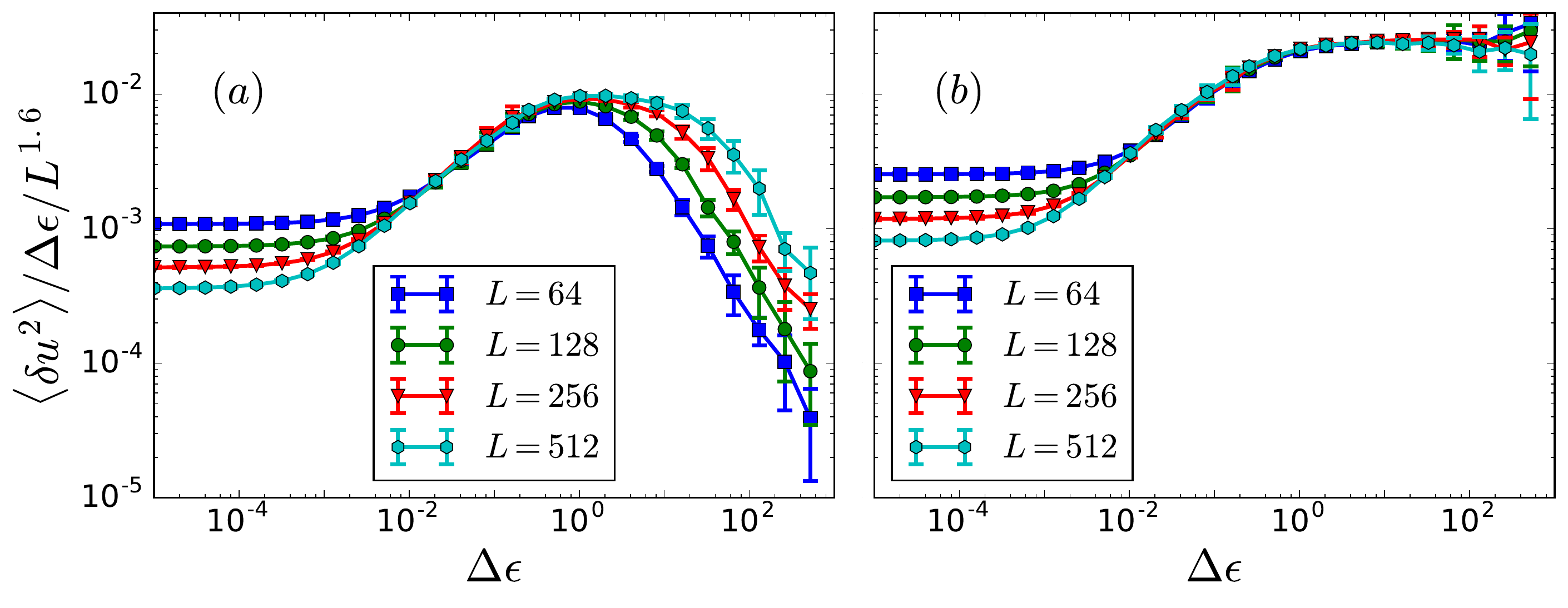}
\caption{\label{fig:diffusion1p6} Diffusivity $D$ scaled by $L^{1.6}$
  as a function of $(\Delta\epsilon)$ for a) Mode~2 and b) Mode~3.}
\end{center}
\end{figure}

In Fig.~\ref{fig:diffusion1p6} we again plot the diffusion coefficient
$D$ vs the strain $\epsilon$, but now with $D$ scaled by
$L^{1.6}$ to collapse the upper plateau at long time.  We see a
crossover to the upper plateau at a strain of order unity regardless
of $L$.  This occurs after the departure of $\langle
\epsilon_p^2\rangle$ from its short time plateau.  The mode~3 case displayed in Fig \ref{fig:diffusion1p6} (b)
remains perfectly diffusive for as long as we can simulate and we have
no reason to believe it will do otherwise.

The mode~2 case shown in Fig \ref{fig:diffusion1p6} (a) shows a strikingly different behavior.  For any finite
$L$, the mean square displacement eventually saturates and $D$
decays like $1/\Delta\epsilon$ at long enough $\Delta\epsilon$.
Despite this decay, we observe the emergence of an \emph{apparent}
upper plateau before the decay even for mode~2 at sufficiently large
$L$.  Furthermore, the height of the plateau seems to obey the
$L^{1.6}$ scaling as well.

The emergence of the upper plateau in mode~2 appears to be
related to the spectral gap in the Eshelby convolution operator
disappearing in the $L\rightarrow\infty$ limit.  For
sufficiently large systems, there will be little difference between
mode2 and mode3 loading, despite the lack of zero modes of the Eshelby
convolution operator in the former, as long as $\Delta\epsilon$
remains below the onset of decay.

\begin{figure}[ht]
\begin{center}

\includegraphics[width=9cm]{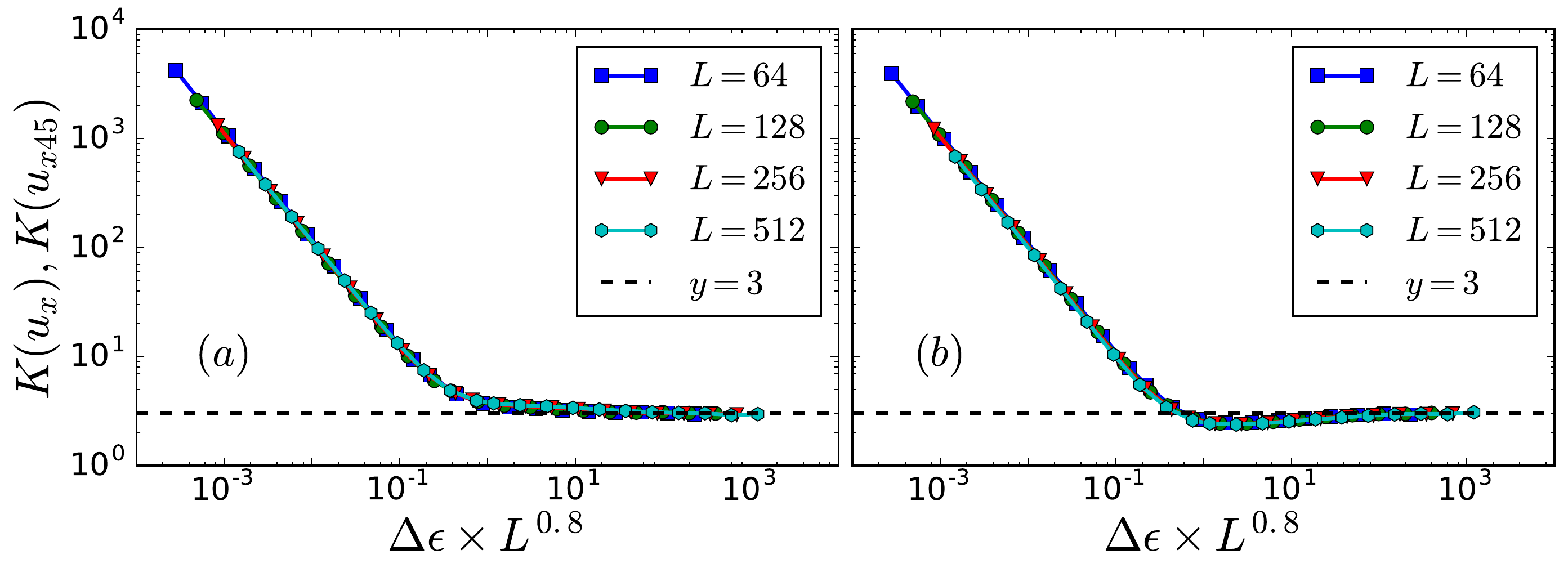}
\caption{\label{fig:kurtosis}Kurtosis of the $P(u_x)$ (a) and $P(\frac{u_x+u_y}{\sqrt{2}})$ (b) distributions for mode 3.
Mode 2 is indistinguishable from mode 3 after interchange of $(\frac{u_x+u_y}{\sqrt{2}})$ and $u_x$.}
\end{center}
\end{figure}
In figure~\ref{fig:kurtosis}, we plot the kurtosis, $K$, of the
distribution of the $x$ (a) and $(x+y)/\sqrt{2}$ (b) Cartesian
component of the displacement versus $\Delta\epsilon$ for
mode~3~\footnote{We find virtually identical behavior for mode2 under
  rotation of the Cartesian components by $45$ degrees.}  Both plots
show a striking initial $K\sim 1/\Delta\epsilon$ behavior as observed
earlier in Lennard-Jones glasses~\cite{Tsamados-PhD09, Tsamados2009a}.
The $K\sim 1/\Delta\epsilon$ behavior can be explained as follows.
Suppose the displacement field is built up from a succession of
characteristic events which are spatially uncorrelated with each
other.  Furthermore, suppose we are interested in a timescale
$\Delta\epsilon$ for which it is unlikely to observe more than one
event in a given window.  Then a typical window of duration
$\Delta\epsilon$ contains either one event (with probability
$\Delta\epsilon/\epsilon_*$) or no event (with probability
$1-\Delta\epsilon/\epsilon_*$) where $\epsilon_*$ is the
characteristic strain release in the event.  So any particular moment
of the distribution (in particular the second and fourth) should scale
like $\langle \delta u^n \rangle\sim\Delta\epsilon$.  So in
particular, for the kurtosis, $\langle \delta u^4 \rangle /\langle
\delta u^2 \rangle^2\sim
\Delta\epsilon/\Delta\epsilon^2=1/\Delta\epsilon$ which is precisely
what we see and explains the much earlier atomistic results from
Tsamados {\it et. al.}\cite{Tsamados-PhD09, Tsamados2009a}: at long times we recover $K \approx 3$, an indication of a Gaussian-like distribution.  We would
expect our data for different system sizes to collapse when rescaled
by the characteristic strain, $\epsilon_*$, which was found above, in
the analysis diffusion coefficient, to scale like $L^{-1.05}$.
However, we find the best collapse when $\Delta\epsilon$ is scaled by
$L^{0.8}$ and the discrepancy between the characteristic strain
inferred from the diffusion coefficient and the kurtosis remains an
outstanding puzzle.

To summarize, we have shown that lattice models for athermal quasistatic amorphous plasticity show good agreement with particle-based simulations for the system size dependence of the short-time diffusion coefficient.
This is for two different stochastic prescriptions for the local energy landscape: random threshold or random plastic strain increment; two different dynamical update rules: synchronous or extremal; and two different orientations of the loading with respect to the lattice.
Our results are also in agreement with Maloney and Robbins~\cite{Maloney2008} who showed that the variance of the local strain field shows little size dependence, while the variance of the displacements shows dramatic size dependence.

At longer times, the diffusion coefficient shows a size dependence,
$D_e\sim L^{1.6}$ which is similar to the $D_e\sim L^{1.5}$ observed
by Martens {et. al.}~\cite{Martens2011a}  In this long time regime, the behavior is
different for the two different modes of loading.  When loading along
the axes of the lattice, the discretized Eshelby kernel has no null
modes, so the variance of the plastic strain, and thus the variance of
the displacements, saturates.  When loading $45$ degrees away, the
Eshelby kernel has proper null modes -- perfect slip lines along the
lattice axes which leave the stress field uniform, so the variance can
continue to grow and the system can achieve a proper diffusive limit
in agreement with earlier arguments by Tyukodi {et. al.}.~\cite{TPRV-PRE16}  We note
that even in the axial-load case where there are no perfect null modes
of the kernel, a pseudo-diffusive-plateau develops at the very latest
times.  The extent of the pseudo-diffusive-plateau depends on system
size with larger sizes maintaining a quasi-diffusive regime for a
longer period of time, but a precise study of the long-time diffusive
behavior is left for future work.

The present picture we put forward here of a separate early time
  diffusive regime crossing over to a distinct late time diffusive
  regime clarifies the apparent discrepancy between particle-based and
  lattice-based models. In particular, it appears that the
  introduction of convection discussed by Nicolas
  et. al.~\cite{Nicolas2014b} is not necessary to recover the linear
  size scaling of the correlations observed in atomistic
  simulations. In light of our present work, it seems likely that a
  $D_e\sim L^1$ regime was already present in former advection-free
  lattice models based studies but that this early difusive regime was
  simply not analyzed. Of course, at very late times, advection should
  be important for a detailed comparison with particle-based
  simulations.

\begin{acknowledgments}
This material is based upon the work supported by the National Science Foundation under Grant No. DMR-1056564 and in part by Grant No. PHY-1748958. 
CEM would like to acknowledge discussions with Kirsten Martens which motivated this study.

\end{acknowledgments}

\bibliography{library}

\end{document}